# High-entropy alloy TiV$_2$ZrCrMnFeNi for hydrogen storage at room temperature with full reversibility and good activation

Shivam Dangwal[1,2] and Kaveh Edalati[1,2,*]

[1] WPI, International Institute for Carbon-Neutral Energy Research (WPI-I2CNER), Kyushu University, Fukuoka 819-0395, Japan
[2] Department of Automotive Science, Graduate School of Integrated Frontier Sciences, Kyushu University, Fukuoka 819-0395, Japan

The development of alloys that are hydrogenated and dehydrogenated quickly and actively at room temperature is a challenge for the safe and compact storage of hydrogen. In this study, a new high-entropy alloy (HEA) with AB-type configuration (A: hydride-forming elements, B: inert-to-hydrogen elements) was designed by considering valence electron concentration, electronegativity difference and atomic-size mismatch of elements. The alloy TiV$_2$ZrCrMnFeNi had dual C14 Laves and BCC phases, in which C14 stored hydrogen and BCC/C14 interphase boundaries contributed to activation. The alloy absorbed 1.6 wt% of hydrogen at room temperature without any activation treatment and exhibited fast kinetics and full reversibility.

*Corresponding author
 Kaveh Edalati (E-mail: kaveh.edalati@kyudai.jp; Tel/Fax: +81 92 802 6744)

The prosperity and well-being of humans are directly linked with the growth in the energy sector. The source of energy can be renewable (e.g. wind, solar, hydrothermal, etc.) and non-renewable (e.g. oil, coal, etc.) [1]. Non-renewable fossil fuels have limited amounts and their impact on the environment is devastating. Therefore, there is a global trend to use renewable sources of energy as well as clean energy carriers that have sufficient efficiency for future demands [1]. A clean energy carrier must be producible in large quantities and its by-products should not be harmful to the environment. Hydrogen satisfies the above conditions as it can be produced efficiently from water and its combustion produces only water [1]. Since hydrogen is the lightest element, its storage is a critical issue [2]. The conventional methods of storing hydrogen are in the form of compressed gas and liquid, but these methods have some drawbacks such as special tank requirements, easy vaporization and safety issues related to high-pressure and cryogenic conditions [2]. A potential alternative is storing hydrogen in the form of metal hydrides [3,4].

Most of the binary metal hydrides are not suitable for hydrogen storage as they require high temperatures or low pressures for dehydrogenation. In 1958, scientists came up with ternary metal hydrides as a new class of hydrides produced from intermetallics containing both A- and B-type elements, where A indicates the elements with a strong affinity for hydrogen and B indicates the elements with a weak affinity for hydrogen. [5]. The thermodynamic stability of these ternary hydrides is lower than corresponding A-type hydrides leading to a decrease in the desorption temperature [3-5]. The seven-decade research on ternary hydrides could introduce only limited materials such as $LaNi_5$ and TiFe that can reversibly store hydrogen at room temperature, but these materials have some other limitations [3]. For example, $LaNi_5$ contains a rare-earth element and has low air resistance [6] and TiFe requires an activation process for storing hydrogen by either high-temperature treatment [7], severe mechanical deformation [8] or chemical modification [9]. The most recent solution to tune the thermodynamic stability of hydrides for hydrogen storage is the use of multi-component high-entropy hydrides which are produced from high-entropy alloys (HEAs) [10], as schematically shown in Fig. 1.

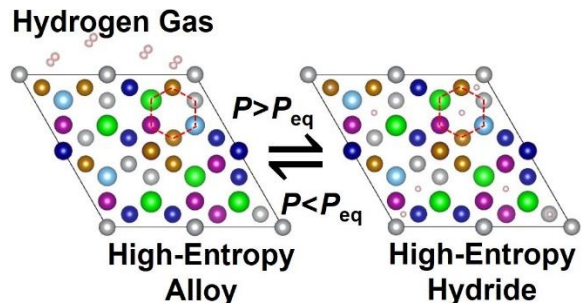

Figure 1. Illustration of hydrogen absorption and desorption in high-entropy alloys and corresponding hydrides ($P_{eq}$: equilibrium plateau pressure).

The concept of HEAs was first put forward in 2004 as alloys containing at least five principal elements with 5-35 at% of each element [11]. Another definition of HEAs is the materials whose configurational entropy is greater than $1.5R$, where $R$ stands for the gas constant [12]. Many HEAs were examined for hydrogen storage within past years [12] such as MgTiVCrFe (desorption temperature > 623 K) [13], TiVCrNbMo (desorption temperature: 523 K) [14], TiVCrNbTa (desorption temperature: 473 K) [14], TiZrHfMoNb (desorption temperature: 575 K) [15], TiZrNbMoV (no desorption temperature reported) [16], TiVZrHfNb (desorption temperature: 593 K) [17] and Ti-V-Zr-Cr-Mn-Fe (activation temperature: 673 K) [18]. However, almost all these alloys either need a high-temperature activation or desorb hydrogen at high temperatures [12]. A recent study used both calculations and experiments and suggested three criteria for designing HEAs for hydrogen storage at room temperature: (i) $AB_2$-type configuration, (ii) C14 Laves phase structure, and (iii) valence electron concentration (VEC) of 6.4 [19]. Using the above approach, an $AB_2$-type HEA TiZrCrMnFeNi was designed, which exhibited 1.7 wt% hydrogen absorption at room temperature [19]. It was later shown by first-principles calculations that these three criteria can lead to a low hydrogen binding energy for room-temperature hydrogen storage [20]. Another two studies suggested that VEC can be in a wider range of 6.2-6.4 and the atomic configuration is not limited to $AB_2$ type [21,22]. Despite these successful attempts at the adjustment of thermodynamic stability of high-entropy hydrides for hydrogen storage, it is still demanding to develop new HEAs that not only store hydrogen at room temperature but also do not need any activation treatment.

The objective of this study is to design a HEA that can absorb and desorb hydrogen at room temperature without the need for activation treatment. To achieve this objective, an AB-type alloy with the C14 Laves phase structure and some amounts of a BCC phase was developed. For the development of the alloy, the $AB_2$-type alloy introduced in the authors' earlier study [19], was used as a master alloy and vanadium as a BCC stabilizer was added to it to achieve an AB-type composition while satisfying three criteria: (i) VEC should be higher than 6 [19-22], but it should be well below 6.87 to avoid the formation of FCC phase [23]; (ii) the electronegativity difference between elements should be higher than 7% and atomic size mismatch should be larger than 5% to form C14 Laves phase [24]; and (iii) small amount of BCC phase should be introduced by increasing the fraction of BCC-stabilizing elements such as vanadium [25] so that C14/BCC interphase boundaries are formed as potential hydrogen pathway for easy activation [26]. Vanadium is selected in this study not only because it is a lightweight BCC stabilizer in HEAs [25] but also because it has a high affinity with hydrogen as reported in T-V-Cr hydrogen storage alloys [27]. $TiV_2ZrCrMnFeNi$ with VEC of 6.125, atomic size mismatch of 8.41%, and electronegativity difference of 10.21% with an extra amount of vanadium was designed which could reversibly store hydrogen at room temperature without the need for any activation treatment.

The HEA $TiV_2ZrCrMnFeNi$ was prepared using vacuum arc melting from high-purity titanium (99.9%), vanadium (99.7%), zirconium (99.2%), chromium (99.99%), manganese (99.9%), iron (99.9%) and nickel (99.99%). The alloy was rotated and re-melted seven times to achieve a homogeneous composition. After arc melting, the ingot was cut into cylindrical discs with a diameter of 10 mm and 0.8 mm thickness using electric discharge machining.

First, the crystal structure of the alloy was examined after crushing by X-ray diffraction (XRD) using Cu Kα radiation with an acceleration voltage of 45 kV and a filament current of 200 mA. To determine the lattice parameters, Rietveld analysis in PDXL software was used.

For microstructural analysis, the discs were ground with emery papers (grit number – 800, 1200, 2000), polished with 9 µm and 3 µm diamond suspensions and finally fine-polished by colloidal silica with 60 nm particle size. The microstructural analysis was done using scanning electron microscopy (SEM) with an acceleration voltage of 15 kV. The energy dispersive X-ray spectroscopy (EDS) was performed for compositional analysis and electron backscatter diffraction (EBSD) was used for phase analysis. The SEM images were also analyzed using the MATLAB image segmentation tool to determine the phase percentages.

To analyze the nanostructure of the alloy, transmission electron microscope (TEM) and scanning-transmission electron microscopy (STEM) were used with an acceleration voltage of 200 kV. Ethanol was used as a medium to crush the alloy while preventing oxidation. After being crushed, it was dispersed onto a carbon grid and immediately examined by high-resolution imaging, fast Fourier transform (FFT), high-angle annular dark-field (HAADF) imaging and EDS mapping.

To examine the hydrogen absorption-desorption properties and kinetics ability of the HEA, a Sievert-type machine was used. The alloy was crushed in air and sieved through a 75 µm sieve. The crushed sample was transferred to the reactor of the Sievert-type equipment and was evacuated at room temperature for 3 h to remove possible moisture. After that, pressure-temperature-composition (PCT) isotherms were determined for 3 cycles at 303 K and a kinetics test was subsequently conducted for 1 h under a hydrogen pressure of 3.5 MPa. The hydrogenated sample after the kinetics test was exposed to air to trap the hydrogen in the hydride form, and its crystal structure was immediately analyzed using XRD. Furthermore, cyclic hydrogen absorption and desorption measurements were performed for 50 cycles (10 min hydrogen absorption under an initial pressure of 3.7 MPa at 303 K followed by 120 min evacuation at 303 K).

The crystal structure analysis using XRD is shown in Fig. 2. The XRD analysis and corresponding Rietveld refinement reveal that the ingot is made up of 2 phases: (i) 84 wt% of C14 Laves phase (P6$_3$/mmc with lattice parameters of $a = b = 0.494$ nm and $c = 0.813$ nm), and (ii) 16 wt% of BCC phase (Im-3m with lattice parameters of $a = b = c = 0.295$ nm). The formation of C14, which is known as an appropriate phase for reversible hydrogen storage in conventional alloys [27,28] and HEAs [18-22], should be due to the adjusted electronegativity difference and atomic-size mismatch [24], while the formation of BCC should be due to the high vanadium concentration [25].

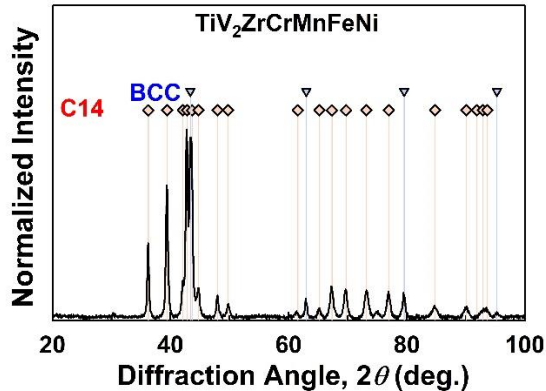

Figure 2. Formation of C14 Laves phase (P63/mmc with $a = b = 0.494$ nm and $c = 0.813$ nm), and BCC phase (Im-3m with $a = b = c = 0.295$ nm) in high-entropy alloy TiV$_2$ZrCrMnFeNi. XRD profile and peak position for C14 and BCC phases.

The microstructure examined by SEM, as shown in Fig. 3(a), reveals the presence of round dark particles in a bright matrix. The image analysis using MATLAB software reveals that the dark particles with an average size of 2.24 µm occupy ~20% of the image whereas the bright matrix with an average grain size of 18.6 µm occupies ~80% of the image. A comparison between these volume fractions with XRD analysis suggests that the dark particles and white matrix should have BCC and C14 structures, respectively. Fig. 3(b) shows the EBSD phase map with the blue color indicating the BCC phase and the red color indicating the C14 Laves phase. EBSD confirms that the BCC phase corresponds to the dark particles whereas the C14 Laves phase corresponds to the bright matrix. The EDS elemental mapping, shown in Fig. 3(c) and quantitively summarized in Table 1, confirms that the BCC phase is rich in vanadium and chromium and almost free from zirconium. The high concentration of vanadium in BCC is reasonable as it is a well-known BCC stabilizer [25] and the high concentration of zirconium in the Laves phase is reasonable because it generates a large atomic-size mismatch to stabilize C14 [24,29].

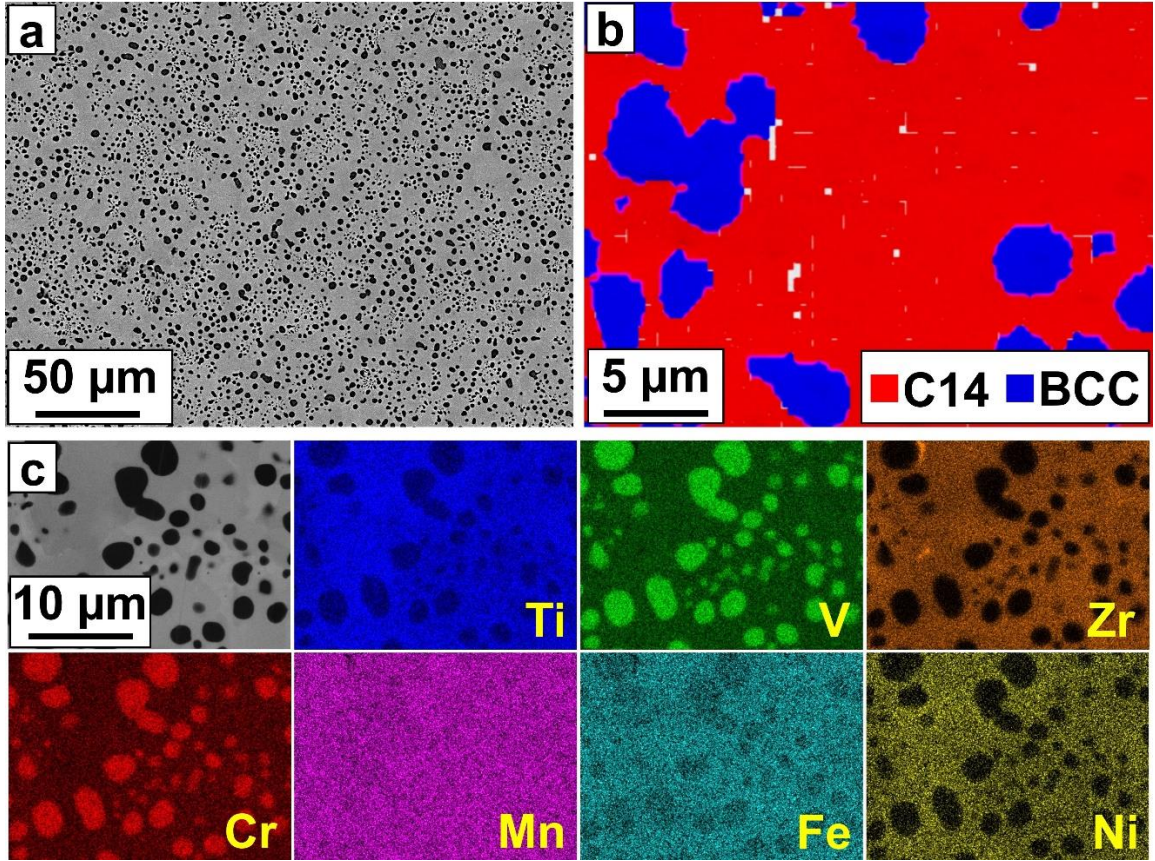

Figure 3. Formation of V- and Cr-rich and Zr-poor BCC phase within C14 Laves phase matrix of high-entropy alloy TiV$_2$ZrCrMnFeNi. (a) SEM micrograph, (b) EBSD phase map with a beam step size of 1 µm, and (c) SEM micrograph and corresponding EDS elemental mappings.

Table 1. Lattice parameters, phase fractions, average grain sizes, and chemical compositions of C14 and BCC phases in high-entropy alloy TiV$_2$ZrCrMnFeNi.

|  | Evaluation Method | Nominal | Overall | C14 Phase | BCC Phase |
|---|---|---|---|---|---|
| **Lattice Parameters (nm)** | XRD |  |  | $a = 0.494$ $c = 0.813$ | $a = b = c$ $= 0.295$ |
| **Fraction (wt%)** | XRD |  |  | 84 | 16 |
| (vol%) | SEM |  |  | 80 | 20 |
| **Grain size** (µm) | SEM |  |  | 8.6 | 2.2 |
| **Ti (at%)** | EDS | 12.5 | 12.8 | 15.3 | 5.4 |
| **V (at%)** |  | 25 | 27.7 | 19.2 | 51.2 |
| **Zr (at%)** |  | 12.5 | 10.9 | 15.2 | 0.0 |
| **Cr (at%)** |  | 12.5 | 13.8 | 10.3 | 23.5 |
| **Mn (at%)** |  | 12.5 | 10.1 | 10.4 | 9.0 |
| **Fe (at%)** |  | 12.5 | 11.9 | 13.7 | 8.4 |
| **Ni (at%)** |  | 12.5 | 12.8 | 15.9 | 2.5 |

Fig. 4(a) shows EDS elemental mappings using STEM which indicates the proper mixing of the elements in the alloy at the nanometer level. The presence of V-rich and Zr-poor in Fig. 4(a) should be due to the presence of the BCC phase, as discussed in Fig. 3. High-resolution analysis, as shown in Figs. 4(b) and 4(c) confirm the co-presence of C14 and BCC phases which is in good agreement with XRD and EBSD analyses. The co-presence of these two phases at the nanometer scale can be beneficial for easy activation because earlier studies showed that grain and interphase boundaries are effective pathways for hydrogen transport [9,26].

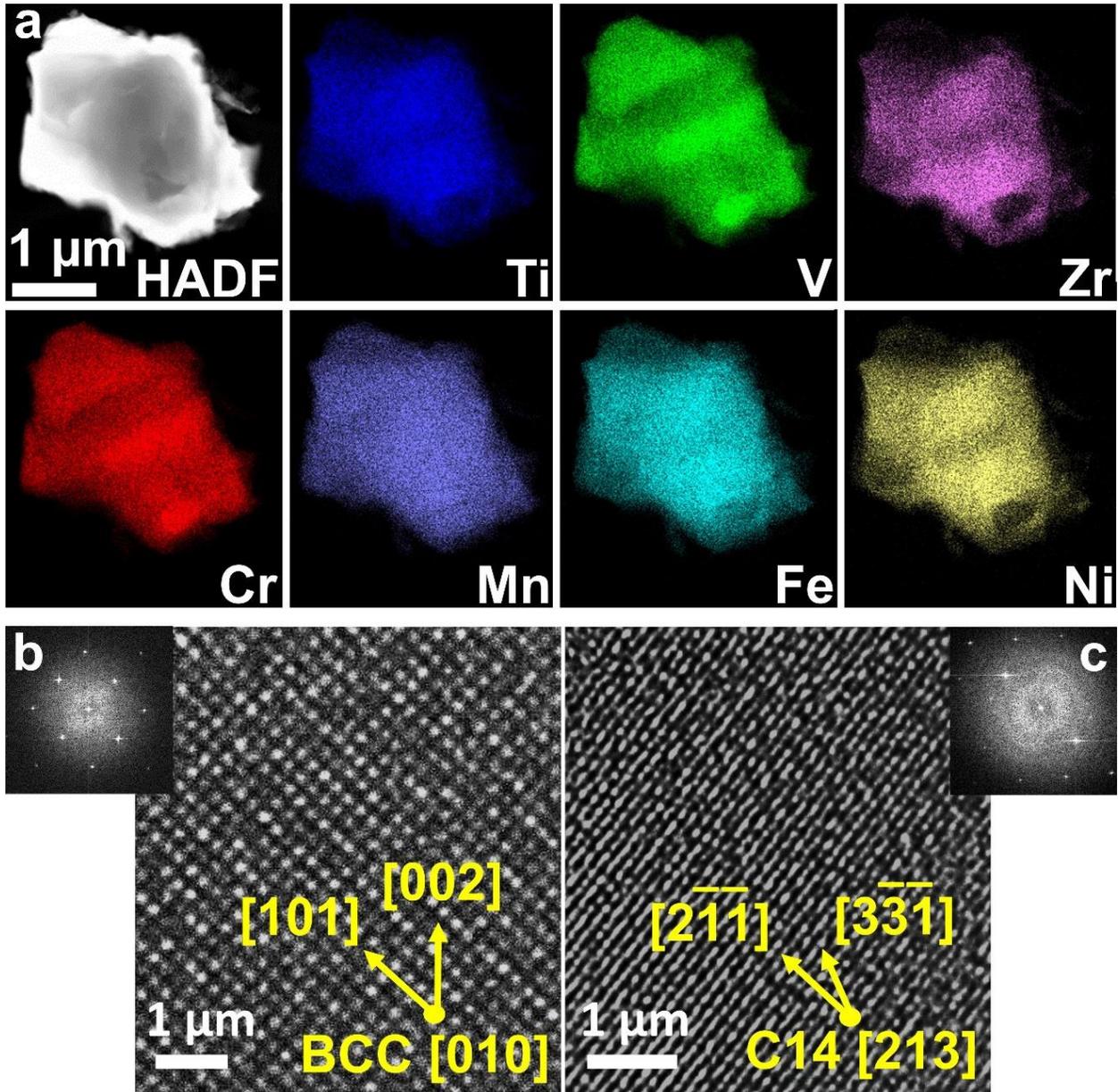

Figure 4. Appropriate mixing of elements at the nanometer level in C14 Laves and BCC phases of high-entropy alloy TiV$_2$ZrCrMnFeNi. (a) STEM-HAADF micrograph and corresponding EDS elemental mappings, and high-resolution TEM lattice images and corresponding FFT diffractograms of (b) BCC and (c) C14 Laves phase.

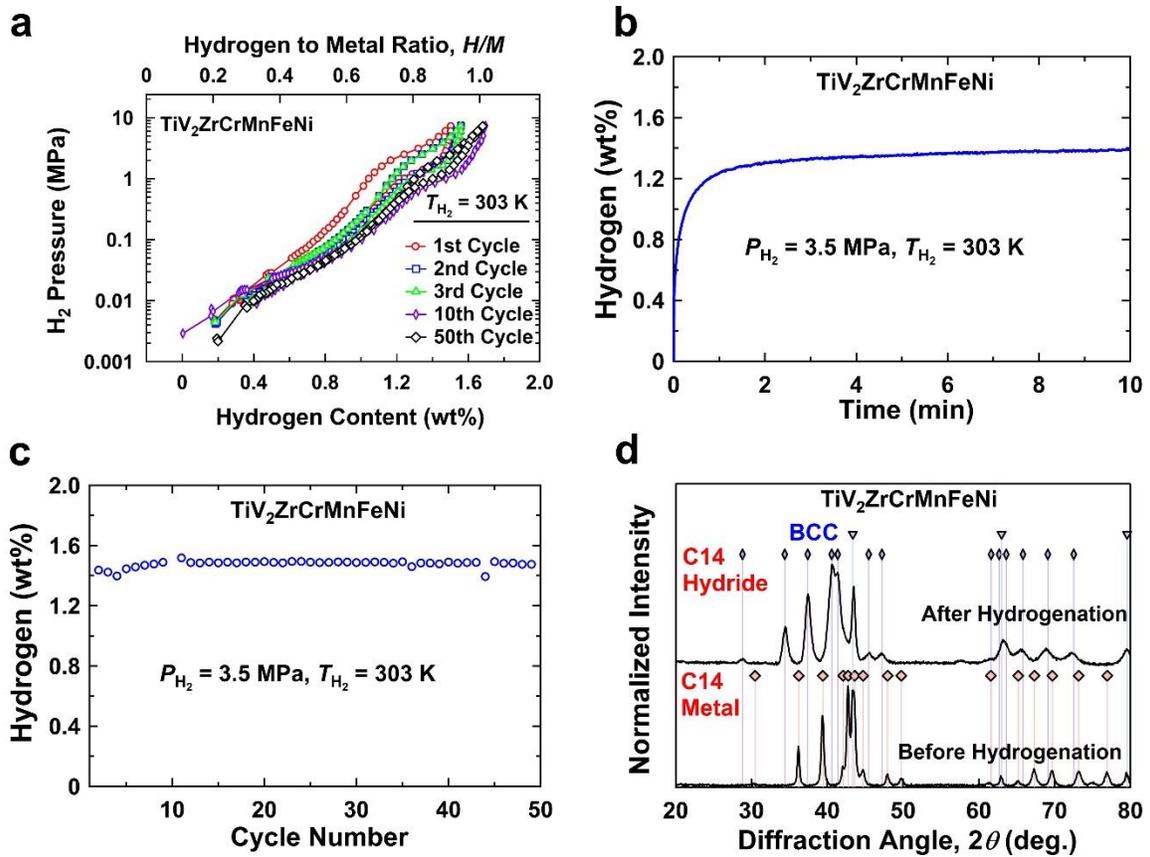

Figure 5. Reversible and quick storage of hydrogen at room temperature in high-entropy alloy TiV$_2$ZrCrMnFeNi in the form of Laves phase hydride ($a = b = 0.519$ nm and $c = 0.849$ nm). (a) PCT absorption/desorption isotherms at 303 K, (b) hydrogenation kinetics curve at 303 K, (c) hydrogenation cycling tests for 50 cycles at 303 K, and (d) XRD profile before and after hydrogenation.

The capability of the HEA to reversibly absorb and desorb hydrogen is shown in Fig. 5(a) using PCT isotherms. The HEA absorbs 1.6 wt% of hydrogen (hydrogen-to-metal atomic ratio: $H/M = 0.92$) at room temperature without the need for any activation process and desorbs the hydrogen completely. This reversible storage capacity does not decrease with cycling for at least three cycles. Moreover, this storage capacity is comparable to LaNi$_5$ and TiFe [7-9], but the current HEA works at room temperature without any activation process and it remains active even after keeping and crushing under an air atmosphere. The kinetics test shown in Fig. 5(b) shows that the material reaches a steady state in less than one minute, indicating very fast kinetics. This fast kinetics should be due to the structural features of the Laves phase [30] but its easy activation should be influenced by the presence of C14/BCC interphases [9,26]. To achieve such fast hydrogenation, using catalysts [3,4] or generating nanostructure by high-energy ball milling [26,31] or severe plastic deformation [32,33] is usually needed. The cyclic hydrogen absorption/desorption test, shown in Fig. 5(c), confirms that the cycling performance of the HEA is acceptable for at least 50 cycles. The crystal structure analyzed by XRD after hydrogenation reveals that the peaks of the C14 Laves phase shift to lower diffraction angles whereas the peaks of the BCC phase remain at the same position, as shown in Fig. 5(d). There is a 15% expansion in the volume of the C14 Laves phase after hydrogenation and the Rietveld analysis suggests that the lattice parameters increase to $a = b = 0.519$ nm and $c = 0.849$ nm. These XRD analyses confirm that hydrogen is stored only in the C14 phase in the form of a Laves phase hydride, while BCC does not contribute as a media for solid-state storage and its lattice parameter remains the same as that before hydrogenation ($a = 0.295$ nm). Since hydrogen atoms occupy the interstitial sites as solute atoms, a clear plateau pressure cannot be detected in Fig. 5(a). Moreover, since hydrogen atoms stay only in the C14 phase and this phase contains 84 wt% of the HEA, it can be concluded that 1.6 wt% of hydrogen corresponds to a metal-to-hydrogen atomic ratio is $H/M = 1.09$ for C14 which is a reasonable ration reported in various kinds of conventional [27-30] and high-entropy [18-22] Laves phase hydrides.

This study not only introduces a new hydride for room-temperature hydrogen storage but also defines some criteria for designing HEAs for room-temperature hydrogen storage. (i) A VEC value higher than 6 [19-22], but well below 6.87 [23] is desirable, while VEC is 6.212 for the C14 Laves phase by considering the composition provided in Table 1. (ii) The C14 Laves phase is considered as the most appropriate phase for room temperature hydrogen storage with a hydrogen-to-metal ratio of ~1 [27-30]. The Laves phase can be stabilized by setting the electronegativity difference and atomic-size mismatch between the elements to over 7% and 5%, respectively [24]. (iii) Although hydrogen storage materials usually are designed to have a single phase to maximize the capacity [13-18], the use of dual-phase alloys is beneficial because the interphase boundaries can act as pathways for hydrogen transport and easy activation [26]. While tuning these three criteria is not easy in binary or ternary systems, HEAs show a high tunability due to the presence of several elements in the solid solution form [11,12]. Tuning these criteria and accordingly weakening the hydrogen binding energy (enthalpy of the formation of hydride) lead to the good performance of current HEA to thermodynamically absorb and desorb hydrogen at room temperature with a metal-to-hydrogen atomic ratio of ~1 [20,34,35]. It is expected that in addition to the effect of interphase boundaries on easy activation, the cocktail effect may positively influence the kinetics by catalytic effect in the HEA with the Laves phase and accordingly enhance the hydrogenation rate through the Jonhson-Mehl-Avrami-Kolomogorov model [36,37]. It should be noted that although earlier studies suggested the significance of interphase boundaries on hydrogenation activation and kinetics of conventional hydrogen storage materials [38-41], future investigations using in-situ synchrotron and neutron diffraction analyses are needed to explain the exact influence of interphase boundaries on hydrogen storage in HEAs. Moreover, although vanadium was used as a lightweight BCC stabilizer [25] in this study, future works with other BCC stabilizers such as iron [42] can further clarify the significance of interphases. Taken all together, dual-phase HEAs show high potential for reversible hydriding and dehydriding at room temperature and this can lead to their future applications not only in hydrogen storage systems but also in nickel-metal hydride (Ni-MH) batteries as they are recent applications of HEAs in Ni-MH batteries [43].

In summary, a dual-phase AB-type HEA $TiV_2ZrCrMnFeNiH_8$ with the majority of C14 Laves phase and a minority of BCC phase was developed for fast and reversible hydrogen storage at room temperature without the need for an activation treatment. While the C14 phase acts as a hydrogen-storing media at room temperature, the BCC phase provides interphases for hydrogen transport and easy activation. The strategy used in this study introduces a possible solution to use dual-phase HEAs as new hydrogen storage materials with appropriate thermodynamics and kinetics at room temperature. Future studies should attempt to develop supplementary strategies to enhance the room-temperature storage capacity from $H/M = 1$ for monohydrides to $H/M = 2$ for dihydrides.

**Acknowledgment**
 The author S.D. thanks the MEXT, Japan for a scholarship. This work is supported in part by grants-in-aid for scientific research from the MEXT, Japan (JP19H05176 & JP21H00150).